\documentstyle[11pt,aasms4]{article}
\def\etal{{\it et al.}}

\def\lap{\hbox{${_{\displaystyle<}\atop^{\displaystyle\sim}}$}}
\def\gap{\hbox{${_{\displaystyle>}\atop^{\displaystyle\sim}}$}}
\def\now{\ifcase\month\or Jan.\or Feb.\or March\or April\or May\or
June\or July\or Aug.\or Sept.\or Oct.\or Nov.\or Dec.\fi
\space\number\day,
\number\year\space (\number\time)}
\def\ky{{\kappa_Y}}

\begin{document}
%
\title{ Observational Constraints on the Internal Structure and
Dynamics of the Vela Pulsar}
%
\author{ Mark Abney,$^{1,2}$ Richard I. Epstein,$^1$ and Angela V.
Olinto$^{1,2}$}
\affil{$^1$Los Alamos National Laboratory,
       Los Alamos, NM 87545, USA}
\affil{$^2$Department of Astronomy and Astrophysics, and Enrico Fermi
Institute,}
\affil{ University of Chicago, 5640 S. Ellis Ave, Chicago, IL 60637}

\begin{abstract}

We show that the short spin-up time observed for the Vela pulsar during the
1988 ``Christmas'' glitch implies that  the coupling
time of the pulsar core to its crust is less than $\sim$ 10
seconds. Ekman pumping cannot explain the fast core-crust
coupling and a more  effective  coupling is necessary. The
internal magnetic field of the Vela pulsar can provide the necessary
coupling if the field threads the core with a magnitude  that exceeds
$ 10^{13}$ Gauss for a normal interior and
$ 10^{11}$ Gauss for a superconducting interior. These lower bounds
favor the hypothesis that the interior of neutron stars contains
superfluid neutrons and protons and challenge the notion that pulsar
magnetic fields decay over million  year time scales or that
magnetic flux is expelled from the core as the star slows.

\end{abstract}

\section{The Vela Christmas Glitch}

Observations of the Vela pulsar during the December 24, 1988
``Christmas'' glitch provide a  glimpse of the coupling  between the
solid crust and the fluid interior of a neutron star. The phase
residuals for this event  are shown in Figure 1 (\cite{MHMK}). The 
nearly linear decay in the  residuals immediately after the glitch shows that
the angular velocity of the surface  first changed  abruptly - within the
$\sim 120$ s time resolution of the observations -  and then varied much more
slowly. As we show below, this abrupt change strongly constrains the internal
structure of the Vela pulsar.

A neutron star is composed of a solid crust containing at most a few
percent of the star's moment of inertia, a liquid core, and a
superfluid neutron liquid that coexists with the crystal  lattice of
the inner part of the crust. If both the neutrons and protons in the
core are superfluid, they  would be strongly coupled by  Fermi liquid
effects (\cite{ALS};  \cite{AS88}). This quantum fluid together with 
the electrons,
 which are needed to maintain charge neutrality, act 
 as a single classical liquid  whose viscosity is supplied mainly
 by the electrons. In this Letter, we study the coupling of this core
liquid  to the star's crust.  

In current models of pulsar glitches (\cite{PA}, \cite{LE96}), the observed
spin up is due to the transfer of angular momentum from the 
neutron superfluid in the inner-crust to the solid crust.
 In these ``crust-initiated" models, subsequent exchange of angular
momentum between the crust and the core  brings these two components
into rotational equilibrium. In alternative ``core-initiated" models,
the initial spin-up occurs in the  core rather than the crust
(\cite{S89}) and the crust then catches up  to the core's angular
velocity.

Models in which glitches are almost entirely a crustal phenomenon,
with only weak coupling to the core, are ruled out by timing data
from accreting neutron stars. This type of ``crust-only" mechanism would
require that the crust of the Vela pulsar remains decoupled from the core for
months, the characteristic post-glitch relaxation time scale.
However,  analyses of timing noise in accretion-powered pulsars 
(Boynton
\& Deeter 1979; Boynton 1981; Boynton \etal\ 1984)  indicate that
$\gap14$\% of the stellar core couples to the crust  in much less
than a month, ruling  out the  possibility of  ``crust-only" glitch
mechanisms.

For either the crust-initiated or core-initiated models, the  Vela
Christmas glitch strongly constrains the core-crust coupling
time-scale. A linear coupling model for crust-interior interaction can 
illustrate the types of constraints different glitch models  yield.  If $I_C$
and $\Omega_C(t)$ are the moment of inertia and  angular velocity of the
solid crust,  $I_I$ and $\Omega_I(t)$ are the moment of inertia and angular
velocity  of the liquid interior (which we take to behave as a solid
body for this schematic example), and $t_{CI}$ is the  coupling time
scale between the crust and the interior, we can write:
\begin{equation} I_C\dot \Omega_C(t) = - \dot J_{CI}(t) + \dot J_{C,
{\rm S}}(t) \ ,
\end{equation}
\begin{equation} I_I\dot \Omega_I (t) =   \dot J_{CI}(t) + \dot
J_{I,{\rm S}}(t) \ , 
\label{eq2}\end{equation}  where
\begin{equation}
 \dot J_{CI}(t) \equiv I_r {\Omega_C(t) -\Omega_I(t) \over t_{CI}},
\end{equation}
 and $$I_r = {I_C I_I\over (I_C+I_I)}$$ is the reduced moment of
inertia. For a crust-initiated glitch the source term  is 
$\dot J_{C,{\rm S}} (t) = \tau_{C}$,  for $ 0<t<t_{su}^C$ 
(and $\dot J_{I,{\rm S}} (t) =0$); whereas for
a core-initiated glitch  $\dot J_{I,{\rm S}}(t) = \tau_{I}$,  
for $ 0<t<t_{su}^I$ (and $\dot J_{C,{\rm S}} (t) =0$), where the torques 
$\tau_{C}$ and $\tau_{I}$ are taken to be constants.

We solve the linear coupling model for both a core- and
crust-initiated glitch and obtain the phase residuals:
\begin{equation}
\Delta \phi(t) = \int_0^t [\Omega_C(0) - \Omega_C(t') ] dt' .
\end{equation}

 For  $t \ge t_{su}^C$ and $t \gg t_{CI}$,
the phase residuals for a crust-initiated glitch are given by, 
\begin{equation}
 \Delta\phi_C (t) =-\frac{\tau_{C}\,
t_{su}^C}{I_C+I_I}\,\left(t-\frac{t_{su}^C}{2}+\frac{I_I}
{I_C}\, t_{CI} \right) \ ,
\end{equation}   while, for  a core-initiated glitch and $t \ge t_{su}^I$,
\begin{equation}
 \Delta\phi_I (t) =-\frac{ \tau_{I} \,
 t_{su}^I}{I_C+I_I}\,\left(t-\frac{t_{su}^I}{2}-t_{CI}
\right)  \ .
\end{equation} 
By fitting the data from the Christmas glitch to the above
equations, taking into account the uncertainty in when the glitch
started, we  constrain the coupling and spin-up times. The short-term
behavior of the glitch model enables us to place upper limits on
these time scales by requiring that the calculated phase residuals agree
with the data to within the observational uncertainty ($\sim 0.2$
milliperiods). Acceptable models lie between the curves shown in
Figure 1, where the dashed curves  show the phase residuals for two
crust-initiated glitch models and the solid curves show results for
two core-initiated models. 

We find that for a crust-initiated glitch
the   core-crust coupling time scale must satisfy
\begin{equation}
 t_{CI} \lap 300 \, {\rm s} \ {I_C \over I_I} \ ,
\end{equation}  and the spin-up time  must be  $ t_{su}^C \lap 1200$
s [this constraint on the spin-up time is in agreement with the
theoretical models of Link and Epstein (1996)].  Since $I_C / I_I
\sim 0.03$, we have 
\begin{equation}
 t_{CI} \lap 10 \, {\rm s} \ . 
\end{equation}  The upper bound on the coupling time scale is much shorter
than the experimental resolution of the spin-rate changes 
due to  the small moment of inertia of the crust. 

Crust-initiated models have shown excellent agreement with the
observed behavior of glitching pulsars (\cite{LE96}) and will be the
main focus of the discussion below. For completeness, we point out
that for core-initiated glitch models the requirement for the
core-crust coupling time scale is less severe, $t_{CI} \lap 440$~s,
and the spin-up time  must satisfy
$t_{su}^I\lap 1200$~s. 

\section{Fluid Dynamics in  Neutron Star Cores}

The requirement that the neutron star core and crust couple in less
than 10 s sets the most stringent constraints yet on the internal
dynamics of these objects. We first review the proposed  mechanisms
for dynamically coupling the components of a neutron star  and then
examine the physics needed to explain the Christmas glitch.

Easson (1979) suggested that Ekman pumping may rapidly bring crust
and core  into corotation.  In  Ekman pumping,   angular momentum is
transferred via an imbalance between pressure and centrifugal forces
in a thin boundary layer. This imbalance drives a circulation from the crust
to the interior.    If this mechanism worked as Easson argued, it would
couple the crust to the interior on time scale
\begin{equation}  t_E \simeq{1 \over 2 \Omega_C E^{1/2}} \simeq 250 {\rm \
seconds\ for \ Vela}\,,
\end{equation} where  $ E=\nu/2 \Omega_C R^2$ is the Ekman number, $R$
is the radius of the star ($R\sim 10^6{\rm cm}$), and $\nu$ is the kinematic
viscosity from Easson and Pethick (1979). If Ekman pumping worked, this
timescale would be too long for crust-initiated glitches but still adequate
for core-initiated glitches. 

Abney and Epstein (1996) reinvestigated  Ekman pumping taking into
account the effects of compressibility and  composition
stratification. They showed that in neutron stars the combined
effects of stratification and compressibility restrict the Ekman
pumping process to a relatively thin zone near the boundary, leaving
much of the interior fluid unaffected and greatly increasing the
spin-up time.

Following Abney and Epstein (1996), we define the dimensionless
``constant-$Y$ compressibility"  as
$\ky\equiv g R / c_{Y}^2$,  where $g$ is the gravitational
acceleration
 and
$c_{Y}$ is the speed that relates the change in pressure, $p$, with
the change in density, $\rho$, when an element of fluid is displaced
while holding a parameter
$Y$ constant, i.e.;
\begin{equation}
\left(\frac{\partial\rho}{\partial p}\right
  )_Y  \equiv\frac{1}{c_{Y}^2} .
\end{equation}
 If the fluid is displaced adiabatically so that the entropy and
composition are fixed, then $c_{Y}$ is the sound speed. For a typical
neutron star,
$g\approx10^{14}{\rm cm/sec^2}$ and
$c_{Y}\approx 10^9{\rm cm/sec}$ (Epstein 1988), therefore, $\ky
\sim 10^2$.

The effects of stratification are characterized by the
Brunt--V\"ais\"al\"a frequency $N$ given by
\begin{equation}
  N^2\equiv g^2\left(\frac{1}{c_{\rm eq}^2}-\frac{1}{c_{Y}^2}\right ),
\end{equation} where $c_{\rm eq}$ which is slightly less than $c_Y$ relates
the change in density to the change in pressure through the equilibrium
stellar interior, \begin{equation}
  \left (\frac{\partial\rho}{\partial p}\right )_{\rm eq}
  \equiv\frac{1}{c_{\rm eq}^2}. \label{ceq1}
\end{equation} For   compressibilities of $\ky \sim 10^2$, Abney
\& Epstein (1996) find that the  Brunt--V\"ais\"al\"a frequency,
$N$
 has to be  $\lap \, 0.2 \, \Omega_C$  for a significant fraction of the
star to undergo Ekman pumping.

The Brunt--V\"ais\"al\"a frequency for the core of neutron stars has
been studied by  Reisenegger and Goldreich (1992)  and  Lee (1995).
The neutron-to-proton  ratio of a fully relaxed neutron star is set
by an equilibrium among strong and weak interactions that minimizes
the free energy. If two mass elements from different depths in the
neutron star are interchanged,   the composition of the mass elements
would revert to those of the local ambient material after a weak interaction
time scale and  there would be no restoring force. 
However, for shorter times the material ``remembers'' its origin (i.e., 
the neutron-to-proton  ratio does not fully adjust) and a restoring force
pulls on the displaced matter. Since the dynamic time scale for
perturbations in a neutron star are  much shorter than the weak-interaction
time scales, the variation of the neutron-to-proton  ratio with depth 
stabilizes radial motions generating a finite Brunt--V\"ais\"al\"a
frequency.   Using the Pandharipande (1971) equation of state, Reisenegger
and Goldreich (1992) estimated

\begin{equation} N \approx 0.05 \left( {\rho \over
\rho_{\rm nuc}}\right)^{1/2} {g \over c_{\rm eq}} 
\end{equation}
 which gives $N
\approx  500 \, s^{-1}$ around nuclear density and
$N \approx 7 \, \Omega_C$ for Vela. Lee (1995) used the Serot (1979)
equation of state to study the radial dependence of $N$, and found a
similar values for $N$ in the outer regions of the core. 

Alternatively, the core of neutron stars may be composed of quark
matter (\cite{IK69}, \cite{I70}, \cite{CN77}, \cite{FM78},
\cite{FJ78}). In this case, we can use the strange matter equation of
state (\cite{W84}, \cite{HZS},
\cite{AFO}) to calculate the Brunt--V\"ais\"al\"a frequency for stars
composed  mostly of quarks. We find that for a gas of massless up and
down quarks and strange quarks with mass $m_s$, 
\begin{equation}
 N \simeq 0.3 {g \over c_{\rm eq}} \left({m_s \over 205 {\rm MeV}}\right)^2
\end{equation}
 for a bag constant $B = (145 {\rm MeV})^4$. The radial dependence of
N is quite mild in this case varying at most by a factor of a few
over 10 km for a 1.4
$M_{\odot}$ quark star. Since
$m_s
\gg 14$ MeV,  $N \gg  0.2 \,
\Omega_C$ for Vela and Ekman pumping is  inhibited.

\section{Vela's Internal Magnetic Field}

 In the absence of Ekman pumping, viscosity couples the  core and
crust on a   time scale
\begin{equation}
 t_{vis} \sim {1\over \Omega_C E}\sim  {\rm\ months}.
\end{equation}  This time  is too long to explain the observed
 behavior during the Christmas glitch and an alternative
mechanism for the core-crust coupling is necessary. Magnetic fields
can  link the  crust to the
core in a time scale comparable to the Alfven travel time through the
core: $t_{CI} \simeq t_A = R / v_A$, where $v_A$ is the Alfven speed.

The relation between the Alfven speed in the interior of neutron stars
and the average magnetic field depends on whether neutrons and protons are 
 normal or superconducting.   In
principle, the quark liquid core could also be superconducting.
Therefore, we 
consider four possibilities  when  estimating the magnetic  coupling time:
(1) the entire liquid core  is normal (i.e.,  the neutron, protons and/or
quark liquids  are  non-superconducting); (2) both the neutrons and protons
(or all the quarks) are superfluid;  (3) the neutrons are superfluid, but
the protons in at least part of the core are normal; and (4) vice-versa. 

In the first case, where the core liquid is everywhere normal, the
coupling time is  $t_{CI} \simeq \sqrt{4
\pi
\rho} R / \bar B$, where $\bar B$ is the average flux density in the
core. The condition that
$t_{CI} \lap 10$ s implies that the magnetic field satisfies
\begin{equation}
\bar B  \gap  10^{13} \left({ \rho \over 10^{15}\; {\rm
g\,cm}^{-3}}\right )^{1/2} 
\left({ R \over 10^6\; {\rm cm}}
\right)
\left({ t_{CI} \over 10 \; {\rm s} }\right)^{-1} \, {\rm Gauss} \ .
\end{equation}   In this case, the   magnetic field threading the
core would have to be stronger than the   surface field, $B_{\rm
surface} \simeq 3.5 \times 10^{12}$ Gauss (\cite{MT77}).

For the second case where all the components of the core are
superfluid,  the magnetic flux is squeezed into flux tubes of
critical field, $H_c
\simeq 10^{15}$ Gauss. In addition, the neutrons and protons are
coupled by Fermi liquid effects (\cite{ALS}, \cite{AS88}).  The increase in
the flux density raises the Alfven speed  by
$(H_c/\bar B )^{1/2}$. The constraint on   the average field in the
core then becomes
\begin{equation} 
\bar B  \gap 10^{11} \left({ \rho \over 10^{15}\; {\rm
g\,cm}^{-3}}\right ) 
\left({ R \over 10^6\; {\rm cm}}
\right)^2
\left({ t_{CI} \over 10 \; {\rm s} }\right)^{-2} \, {\rm Gauss} \
\end{equation} which is  not larger that the surface field.

  In the third case, the normal protons would quickly come into rotational
equilibrium with the crust whereas the superfluid neutrons would take more
than 10 minutes to couple (\cite{SSS}, \cite{F71}).  
In the opposite case (4),  the superfluid protons couple quickly via the
magnetic field while the normal neutrons couple through their interactions with
electrons. Assuming that  the coupling between protons and neutrons is rapid,
then eq. (17) also holds in this case.

\section{Conclusions}

 We have argued that fluid dynamic
processes such as Ekman pumping do not provide the adequate
core-crust coupling that is implied by the
Christmas glitch.  The needed coupling can be provided by a
magnetic field that threads both the core and crust of the star.
If both the neutrons and protons in the core are normal, the required
average radial flux density in the core is $\gap  10^{13}$
Gauss, about three times greater than the surface field deduced
from the spin down rate.  Such a strong internal
magnetic field argues against the core 
containing only normal neutrons and protons. However, if the core did possess
fields of this strength,    the surface field might actually
{\it grow} as the interior field diffuses out through the
crust.  

It is likely that  both the neutrons and
protons are superfluid and the threading field need only have a mean
flux density of
$\gap 10^{11}$ Gauss.  This constraint allows some of the surface magnetic
flux to be confined to the outer crust of the star from which it
could  decay in
$10^6 -10^7$ years. However, since the magnetic diffusion time for
fields that thread the core is
$\gap 10^{10}$ years (\cite{UVR}), the
decay of the crustal magnetic field would  still leave a field of 
$\gap 10^{11}$ Gauss at the surface.

It has been suggested that the magnetic flux tubes in
the superfluid core may be swept out by the
expanding neutron vortex lattice as the rotation of the star
decreases (\cite {DCC}, \cite{SBMT}). Our
constraints on the present internal magnetic field of  the Vela
pulsar indicate that the flux tubes and vortex lines
manage to pass through each other. Since the
present internal magnetic field is
$\gap 10^{11}$ G,  the
initial field would have to be  $B_{\rm initial}  \gap 9 \times  
10^{12}(P_{\rm initial}/1\; {\rm ms})^{-1}$ Gauss for the sweeping interpretation
to be correct. If the initial spin period of the Vela pulsar was $\sim 1$ ms,
the flux sweeping hypothesis would require that the initial internal magnetic
field  be considerably stronger than the present surface field.

Future observations may  provide deeper insights into  the nature of neutron
star interiors. Monitoring of the Vela pulsar  at higher time resolution could
set more stringent constraints on the internal magnetic field.
In addition,  XTE observations of accreting neutron stars can complement
glitch observations in  the study of neutron star interiors.

We would like to thank G. Baym, B. Link, and J. Sauls for helpful
discussions. M.A. and A.O. were supported in part by the DOE at Chicago, by
NASA grant NAG 5-2788 at Fermi National Laboratory, and through a
collaborative research grant from IGPP/LANL. This work was
carried out under the auspices of the U.S. Department of Energy.

\def\nature{{\rm Nature}}
\def\nucphys{{\rm Nuc. Phys.}}
\def\nucphysa{{\rm Nuc. Phys. A}}
\def\physletb{{\rm Phys. Lett. B}}
\def\physrevc{{\rm Phys. Rev. C}}
\def\physrevd{{\rm Phys. Rev. D}}
\def\sovphysjetp{{\rm Soviet~Phys.~JETP}}
\def\ptpl{{\rm Progr.Theor.Phys.Lett}}
\def\ptps{{\rm Prog.Theor.Phys.Suppl.}}
\def\ptp{{\rm Prog. Theor. Phys.}}

\clearpage

\begin{figure}

\caption{Four model fits that delineate the range of acceptable
parameters: dashed lines for crust-initiated models and solid lines for
core-initiated models. Data  for the Vela  Christmas glitch are shown.}

\end{figure}


\begin{thebibliography}{}

\bibitem[Abney \& Epstein 1996]{AE} Abney, M., and Epstein, R.
I.,1996, J. Fluid Mech.,  312, 327 

\bibitem[Ainsworth, Pines,  \& Wambach 1989]{APW} Ainsworth, T., Pines, D.,
\& Wambach, J. 1989, \physletb 222, 173 

\bibitem[Alcock, Farhi, \& Olinto 1986]{AFO} Alcock, C., Farhi, E., \&
Olinto, A. V. 1986, \apj, 310, 261

\bibitem[Alpar, Langer \& Sauls 1984] {ALS} Alpar, M. A., Langer, S.
A., \&\ Sauls, J. A. 1984, \apj, 282, 533

\bibitem[Alpar  \& Sauls 1988]{AS88} Alpar, M. A. \&\ Sauls, J. A.
1988, \apj, 327, 723

\bibitem[Boynton 1981]{B81} Boynton, P. E. 1981, in { IAU
Symposium 95, Pulsars}, ed. R. Wielebinski \&\ W. Sieber (Dordrecht:
Reidel), p. 279

\bibitem[Boynton  \&\ Deeter  1979]{BD79} Boynton, P. E. \&\ Deeter,
J. E. 1979, in { Compact Galactic X-Ray
     Sources}, ed. F. K. Lamb \&\ D. Pines (Urbana: University of
Illinois),  p. 168

\bibitem[Boynton \etal\ 1984]{B84}   Boynton, P. E., Deeter, J. E.,
 Lamb, F. K., Zylstra, G., Pravdo, S. H., White, N. E., Wood, K. S.,
\& Yentis, D. J. 1984, \apj, 283, L53

\bibitem[Chapline \& Nauenberg 1977]{CN77} Chapline, G., \&
Nauenberg, M. 1977, \physrevd, 16, 450

\bibitem[Ding, Cheng \& Chau 1992]{DCC} Ding, K.
Y., Cheng, K. S.,  and Chau, H. F.  1992, in { Isolated Pulsars},
Eds. K. A. Van Riper, R. I. Epstein, \&\  C. Ho (Cambridge:
Cambridge), p42.
                                                              

\bibitem[Easson,  \&\ Pethick  1979]{EP} Easson, I. \&\
Pethick, C. J. 1979, \apj, 227, 995

\bibitem[Easson, I.  1979]{E79} Easson, I.  1979, \apj, 228, 257

\bibitem[Epstein 1988]{E88} Epstein, R. I. 1988, \apj, 333, 880


\bibitem[Fechner \& Joss 1978]{FJ78} Fechner, W. B., \& Joss, P. C.
1978,
\nature, 274, 347

\bibitem[Feibelman 1971]{F71} Feibelman, P. 1971, Phys. Rev. D. 4, 1589

\bibitem[Freedman \& McLerran 1978]{FM78} Freedman, B., \& McLerran,
L. 1978,
\physrevd, 17, 1109

\bibitem[Haensel, Zdunik, \& Schaeffer 1986]{HZS} Haensel, P.,
Zdunik, J. L., \& Schaeffer, R.  1986, Astron. Ap., 287, 244

\bibitem[Itoh 1970]{I70} Itoh, N. 1970, \ptp, 44, 291

\bibitem[Ivanenko \& Kurdgelaidze 1969]{IK69} Ivanenko, D., \&
Kurdgelaidze, D. F.  1969, Lett. Nuovo Cimento, 2, 13

\bibitem[Lee 1995]{L95}Lee, U.1995 {Astron. Astrophys}, 303, 515

\bibitem[Link  \&\  Epstein  1996]{LE96} Link, B. \&\  Epstein, R. I.
1996, \apj,457,844

\bibitem[Manchester and Taylor 1977]{MT77}Manchester, R. N.  and
Taylor, J. H. 1977, Pulsars (Freeman, San Francisco)

\bibitem[McCulloch \etal\ 1990]{MHMK} McCulloch, P. M., Hamilton, P.
A., McConnell, D., \& King, E. A. 1990, \nature, 346, 822

\bibitem[McKenna  \&\ Lyne 1990]{ML90} McKenna, J. \&\ Lyne, A. G.
1990, \nature, 343, 349

\bibitem[Pandharipande 1971]{P71} Pandharipande, V. R.  1971,
\nucphysa, 178, 123

\bibitem[Pines \& Alpar 1985]{PA} Pines, D. \& Alpar, M. A. 1985,
Nature, 316, 27 

\bibitem[Reisenegger \& Goldreich 1992]{RG} Reisenegger, A. \&
Goldreich, P. 1992, \apj 395, 240

\bibitem[Sauls, Stein, \& Serene 1982]{SSS} Sauls, J. A.,  Stein, D. L., \&
Serene, J. W. 1982, Phys. Rev. D. 25, 697

\bibitem[Sauls 1989]{S89} Sauls, J. A. 1989 in  Timing Neutron Stars,
ed. H. Ogelman \& E.P.J. van den Heuvel (Dorchester: Kluwer), 48

\bibitem[Serot 1979]{S79} Serot, B. D. 1979, \physletb, 86, 146


\bibitem[Srinivasan \etal\ 1990]{SBMT} Srinivasan, G., Bhattacharya,
D. Muslimov, A. G., and Tsygan, A. I. 1990, {\rm Current Sci.}, 
59, 31.

\bibitem[Takatsuka  \& Tamagaki 1995]{TT95}Takatsuka, T.  \&
Tamagaki, R. 1995,
\ptpl, 94, 457

\bibitem[Urpin \& Van Riper 1993]{UVR} Urpin, V. A.,  \& Van Riper,
K. A. 1993, \apj 411, L87

\bibitem[Van Riper, Link, \& Epstein 1995]{VLE}Van Riper, K. A.,
Link, B., \& Epstein, R, I. 1995, \apj, 448, 294


\bibitem[Witten 1984]{W84} Witten, E. 1984, \physrevd, 30, 272


\end{thebibliography}
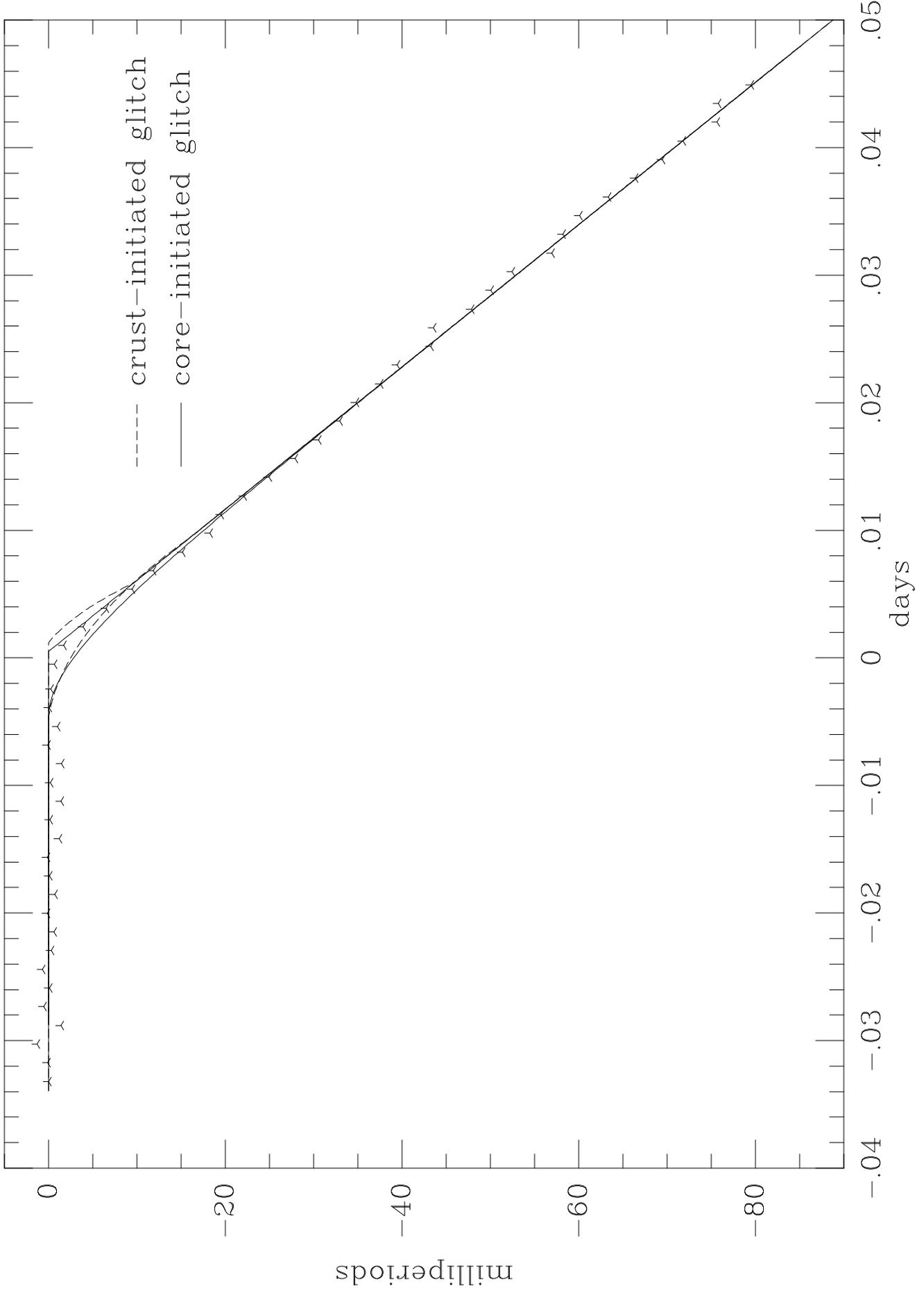
\end{document}